\documentclass[prl,aps,showpacs,twocolumn,preprintnumbers,
amsmath,amssymb,superscriptaddress,longbibliography]{revtex4-1}
\usepackage[page]{appendix} 
\usepackage[english]{babel}
\usepackage{amsmath,amssymb,amsfonts}
\usepackage{graphicx}
\usepackage[colorlinks=True,linkcolor=red,citecolor=blue,urlcolor=blue]{hyperref}
\usepackage{bookmark}
\usepackage[dvipsnames]{xcolor} 
\usepackage{braket}
\usepackage{nicefrac}
\usepackage{bm}
\usepackage{bbm}
\usepackage{braket}
\usepackage{slashed}
\usepackage[normalem]{ulem}
\usepackage{array}
\usepackage{makecell}
\newcolumntype{C}{>{$}c<{$}}

\def\ichi{[\chi]}

\renewcommand{\i}{\mathrm{i}}

\newcommand{\ave}[1]{\langle #1 \rangle}

\allowdisplaybreaks
\DeclareMathAlphabet{\zc}{OT1}{pzc}{m}{it}

  \DeclareUnicodeCharacter{2212}{-}

\begin{document}

\title{Transfer learning from Hermitian to non-Hermitian quantum many-body physics}
\author{Sharareh Sayyad}
\email{sharareh.sayyad@mpl.mpg.de}
\affiliation{Max Planck Institute for the Science of Light, Staudtstra\ss e 2, 91058 Erlangen, Germany}
\author{Jose L. Lado}
\affiliation{Department of Applied Physics, Aalto University, FI-00076 Aalto, Espoo, Finland}

\begin{abstract}
Identifying phase boundaries of interacting systems is one of the key steps to understanding quantum many-body models. 
The development of various numerical and analytical methods has allowed exploring the phase diagrams of many Hermitian interacting systems.
However, numerical challenges and scarcity of analytical solutions hinder obtaining phase boundaries in non-Hermitian many-body models. 
Recent machine learning methods have emerged as a potential strategy to learn
phase boundaries from various observables without having access
to the full many-body wavefunction. Here, we show that a machine learning methodology trained solely
on Hermitian correlation functions allows identifying phase boundaries
of non-Hermitian interacting models. 
These results demonstrate that Hermitian machine learning algorithms can be redeployed to non-Hermitian models without requiring
further training to reveal non-Hermitian phase diagrams. 
Our findings establish transfer learning as a versatile strategy to leverage Hermitian physics
to machine learning non-Hermitian phenomena.
\end{abstract}

\maketitle

\paragraph*{ Introduction.}

The interplay between various degrees of freedom in many-body systems results in the emergence of novel phases of matter, including superconducting~\cite{Dos2007, Proust2019, Andrei2020, Sayyad2020, Nomura2022, Kitatani2023}, Mott insulating~\cite{Sayyad2016, Seo2019, Chatzieleftheriou2023, Kim2023, Tzeng2023}, nematic~\cite{Fernandes2014, Samajdar2021, Sayyad2023b, Mukasa2023, Jiang2023} and topological~\cite{Sheng2011, Neupert2011, Pollmann2012, Bauer2013, Del2023, Kim2023f} phases. Due to their inherent complexity, these systems are often studied computationally, using, e.g., quantum Monte Carlo methods~\cite{Troyer2005, Gezerlis2013, Vaezi2021} and tensor network approaches~\cite{Orus2014, Orus2019, Cirac2021}. In recent years, machine learning methods~\cite{Mehta2019, Carleo2019} have 
provided a complementary strategy to rationalize phases of matter, often in combination with conventional quantum many-body methods. 
The demonstrations
of these roles played by machine learning methods in tackling many-body problems results in characterizing different phases of matter~\cite{Broecker2017, Rodriguez2019, Scheurer2020, Aikebaier2022,vanNieuwenburg2017,PhysRevResearch.4.033223,PhysRevLett.120.176401,Greplova2020,PhysRevApplied.20.024054,Tibaldi2023}, deep learning of the quantum dynamics~\cite{Hartmann2019,PhysRevB.98.060301, Reh2021, Mohseni2022},
obtaining many-body wave functions~\cite{Carleo2017, Szabo2020,PhysRevResearch.4.L012010, Glielmo2020, Reh2023}, and optimizing the performance of computational simulations~\cite{Czischek2018}.

\begin{figure}
    \centering
    \includegraphics[width=\columnwidth]{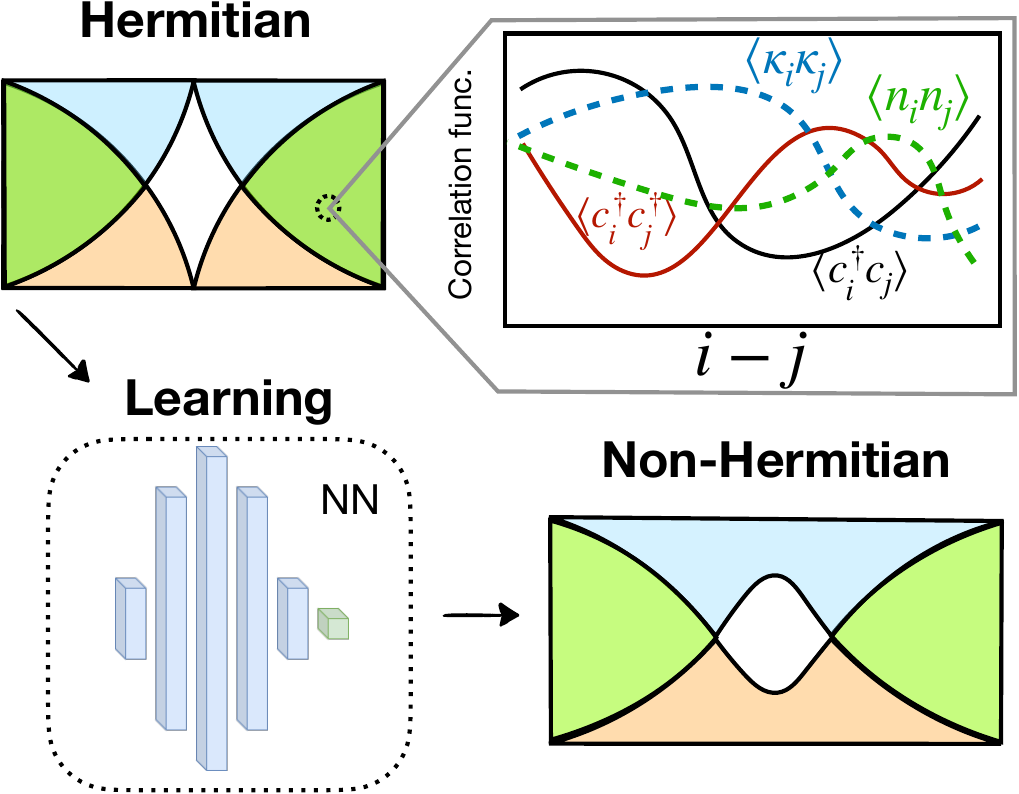}
    \caption{
    \textbf{Non-Hermitian transfer learning:}
    Schematic illustration of the transfer learning methodology
    from Hermitian models to non-Hermitian physics. As an input, for each point of the phase diagram of the Hermitian model,
    short-range two-point~(solid lines) and four-point~(dashed lines) correlation functions are computed (Eqs.\eqref{eq:2points} and \eqref{eq:4points}). The generated correlators for Hermitian systems
    are used to train a machine learning architecture, which in turn
    allows predicting the phase diagram from short-range correlators of the non-Hermitian model.
    The machine learning methodology allows extracting quasi-degeneracies and correlation entropies from the short-range correlators of the non-Hermitian model.
    }
    \label{fig:schematic}
\end{figure}

Exploring correlated physics in open quantum systems attracts great interest mainly because of the systematic treatment of loss and gain in these systems, which quantitatively reproduces experimental observations\cite{Rotter2015, Zhang2022v, PerinaJr2022, Bu2023, Szankowski2023}. In recent years, along with brute force studies of open quantum systems, understanding their effective descriptions based on non-Hermitian physics get momentum~\cite{Ashida2020, Bergholtz2021, Maraviglia2022, Linpeng2022, Banerjee2023, Brzezicki2023}.
The studies of non-Hermitian models have initially focused on single-particle models~\cite{Joglekar2014, Agarwal2015, Agarwal2018, Gong2018, Kawabata2019, Li2019, Wang2020, Sayyad2021, Chen2021, Sayyad2022, Chen2022, Abbasi2022, Sayyad2022b, Kawabata2022, Sayyad2022c, Xiao2022, Kawabata2023}, and its extension to the fully interacting realm has also gained attention recently~\cite{Fukui1998,  Buca2020,  Yamamoto2019, Buca2020, Nakagawa2020, Nakagawa2021, Zhang2021, Nakagawa2021,  Hyart2022, Yoshida2022, Yamamoto2022, Wang2022, Shen2022, Yamamoto2023, Sayyad2023, Shen2023, Han2023, Sayyad2023c, Ghosh2023}. 
Aside from these case studies, unraveling the physics of interacting non-Hermitian systems remains an open challenge due to the scarcity of exactly solvable models, and as conventional (Hermitian) many-body methods cannot be directly applied to the non-Hermitian limit. 
Specifically, obtaining the phase boundaries, understanding the stability of certain phases against non-Hermiticity, and characterizing exotic phases with no Hermitian counterparts remain in general open problems.
Similar to the realm of Hermitian physics, machine learning methods, and specifically
supervised~\cite{Cheng2021, Zhang2021cc, Narayan2021, Ahmed2023}, unsupervised~\cite{Yu2022, Ahmed2023}, and graph-informed methods~\cite{Shang2022} allowed to identify various phases of non-Hermitian non-interacting systems. In these methodologies, the inputs to train learning models are collected from non-Hermitian noninteracting systems and are used to characterize non-Hermitian phase diagrams. As computational methods for Hermitian interacting models are numerically less demanding and more stable than their non-Hermitian counterparts,
learning phase diagrams of non-Hermitian many-body systems from Hermitian correlated models would open up a promising strategy
to leverage many-body methods developed for interacting Hermitian models.

\begin{figure}[t!]
    \centering
    \includegraphics[width=\columnwidth]{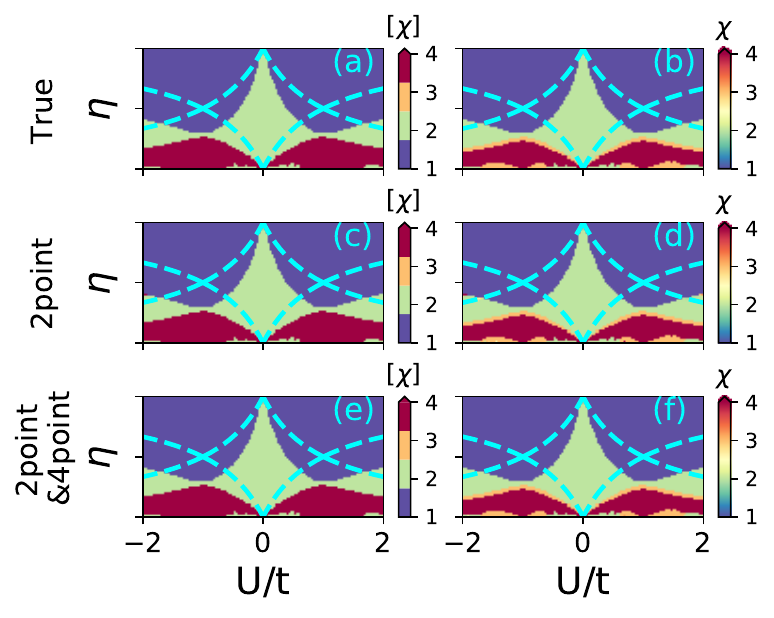}
    \caption{
    \textbf{Hermitian interacting model:}
    The phase diagram of the Hermitian many-body model with $L=16$ on the $U/t-\eta$ plane at $\delta=0.0$. 
    The results in (a) and (b) are calculated by exact diagonalization. Panels (c,d) use a machine learning architecture
     that uses solely two-point correlation functions as input.
     In contrast, panels (e,f) use an architecture trained on both two-point and four-point correlation functions. 
     The quasi-degeneracy in (c,e) is treated as a discrete classifier for $\ichi$, while it is treated as a regression problem in (d,f).
    The boundaries in the thermodynamic limit are shown by cyan dashed lines.
    }
    \label{fig:chiherm}
\end{figure}

In this manuscript, we show that machine learning methods purely trained on Hermitian many-body data can predict interacting regimes in
non-Hermitian interacting models.
For concreteness purposes, we explore the different regimes of the non-Hermitian dimerized Kitaev-Hubbard chain using machine learning techniques schematically shown in Fig.~\ref{fig:schematic}. Here, we collect various correlation functions, 
orders of quasi-degeneracies, and correlation entropies at different parameter regimes of the Hermitian limit of our model.
Using this input, we demonstrate that non-Hermitian regime crossovers can be identified using a machine-learning methodology
trained on short-range Hermitian correlation functions. 
The outcomes of these supervised learning schemes are degrees of quasi-degeneracies and correlation entropies, which can characterize various
regimes of the non-Hermitian model. Our findings reveal that employing correlation entropy as a classifier allows characterizing all regimes of the system.
Our machine-learning approach reliably learns various regimes that share similarities with the Hermitian model. Our method also successfully delineates the regime crossovers even when the correlation effect in the non-Hermitian interacting model deforms the Hermitian phase diagram.

\begin{figure}[t!]
    \centering
    \includegraphics[width=\columnwidth]{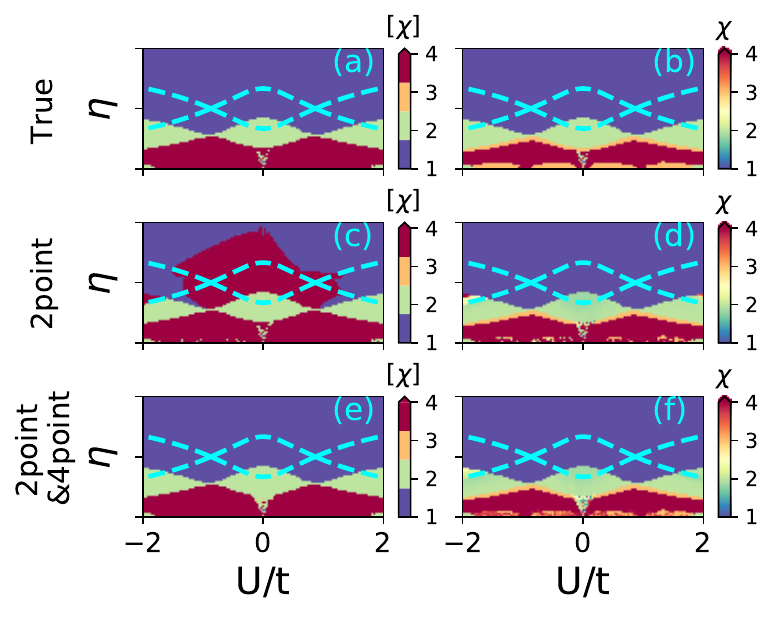}
    \caption{
    \textbf{Non-Hermitian interacting model}:
    The regimes of the non-Hermitian many-body model with $L=16$ on the $U/t-\eta$ plane at $\delta=0.5$. 
    The results in (a) and (b) are calculated by exact diagonalization. The regimes in (c,d) are obtained using architectures
    trained by two-point correlations, whereas (e,f) are trained on both two-point and four-point correlation functions.
    The quasi-degeneracy in (c,e) is treated as a discrete classifier for $\ichi$, while it is treated as a regression problem in (d,f). The boundaries in the thermodynamic limit are shown by cyan dashed lines and black dashed-dotted lines.
    It is observed that while two-point correlators fail to predict the non-Hermitian
    regimes in (c), the inclusion of four-point correlators recovers accurate regime crossovers (e).
    }
    \label{fig:chinonherm}
\end{figure}

\paragraph*{Non-Hermitian interacting model.}
We focus on an interacting non-Hermitian model whose phase boundaries
can be solved exactly in the thermodynamic limit~\cite{Sayyad2023}.
The non-Hermitian dimerized Kitaev-Hubbard Hamiltonian on a chain with length $L$ is given by
\begin{align}
  {\cal H} =&
        - \sum_{j=1}^{L-1}
        \left[ 
        t_{j} \left(   c^{\dagger}_{j} c_{j+1} + 
     c^{\dagger}_{j+1} c_{j} \right)
     +  \Delta_{j}  \left( c^{\dagger}_{j}c^{\dagger}_{j+1} +  c_{j+1}c_{j}  \right)
     \right]
         \nonumber \\
    &
    + \sum_{j=1}^{L-1} (U_{j} - \i \delta_{j} ) \left( 2n_{j} -1 \right) \left(2 n_{j+1} -1 \right)
     ,
     \label{eq:Hintsshk}
\end{align}
where $c^{\dagger}_{j}~(c_{j})$ is a creation~(annihilation) operator for spinless fermion at site $j$ associated with the fermion density $n_{j}=c^{\dagger}_{j} c_{j}$.
Here $t_{j}$, $\Delta_{j}$, and $U_{j}-i \delta_{j}$ denote, respectively, real-valued dimerized hopping amplitude, superconducting pairing amplitude, and complex-valued Hubbard interaction strength. 
Considering the site-independent parameter ${\cal O} \in \{t, \Delta, U, \delta \}$, ${\cal O}_{j} \in \{ t_{j}, \Delta_{j},  U_{j}, \delta_{j} \}$ for $1 \leq j \leq L$ reads
$
    {\cal O}_{j} = 
    {\cal O} (1- \eta) $ 
    if 
    $ j \,{\rm mod}\,2 = 0$
    and ${\cal O}_{j} = 
    {\cal O} (1+ \eta)$ if $j \, {\rm mod}\,2 = 1
$, 
where $\eta$ is the real-valued dimerization parameter.

The Hamiltonian in Eq.~\eqref{eq:Hintsshk} is exactly solvable when $\Delta=t$. At this parameter regime, the interacting model can be mapped to a quadratic fermionic model upon successive two Jordan-Wigner transformations and a spin rotation~\cite{Sayyad2023}. 
Through this procedure, one can show that the spectrum of the effective quadratic Hamiltonian undergoes gap closure upon setting 
$
 \frac{U}{t}=  \pm  \sqrt{\left| \frac{\delta^2}{t^2} - \frac{(1 \pm \eta)^2}{(1 \mp \eta)^2 } \right|},
    \label{eq:phasecond_re}
$ and $
    \frac{U}{t} =
    \pm \frac{1\pm \eta}{ 1 \mp \eta}.
    \label{eq:phasecond_im}
$
These relations ensure the closure of the real-line gaps and the appearance of zero degeneracies in the imaginary part of the spectrum, respectively. Note that these two equations coincide when the non-Hermiticity parameter vanishes, i.e., $\delta=0$.

\begin{figure}
    \centering
    \includegraphics[width=\columnwidth]{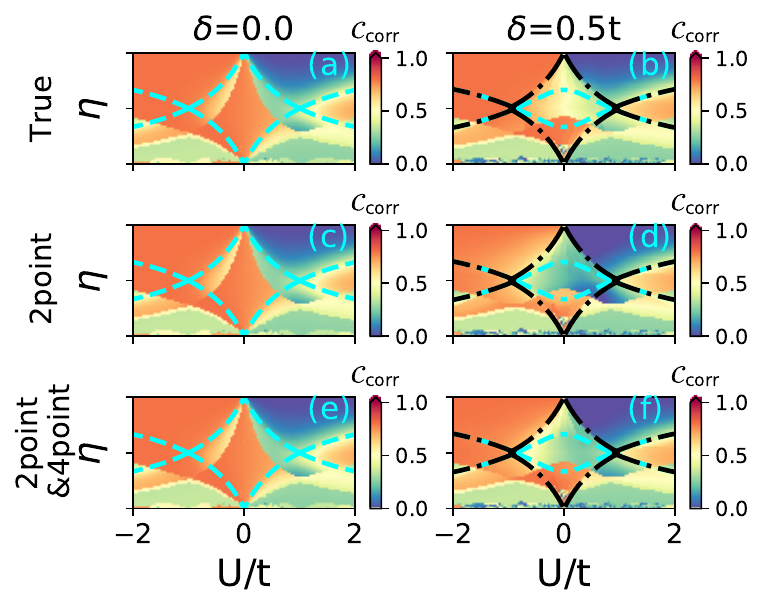}
    \caption{ 
    \textbf{Correlation entropy predictions}: The regimes of the non-Hermitian many-body model with $L=16$ on the $U/t-\eta$ plane at $\delta=0.0$~(a,c,e), $0.5t$~(b,d,f). 
    The trained models are obtained using the Hermitian datasets with $\delta=0.0$. 
    The color bar denotes ${\cal C}_{\rm corr}$. 
    The regimes panels (c,d) are obtained using the machine learning
    model trained by two-point correlation functions,
    whereas (e,f) are trained on both two-point and four-point
    correlation functions. 
    The boundaries in the thermodynamic limit given in the main text
    are shown by cyan dashed lines and black dashed-dotted lines.}
    \label{fig:corrent}
\end{figure}

As $\cal H$ respects the charge conjugation symmetry, eigenvalues come in pairs such that the set of all energies satisfy $\{\varepsilon\}= \{\varepsilon^{*} \}$. This implies that degeneracies of phases can be merely obtained by vanishing real parts of the spectrum.
In a finite system, finite size effects will give rise to small splitting between degenerate states in the
thermodynamic limit.
For finite models, it is thus convenient to define the quasi-degeneracy $\chi$ given by
\begin{equation}
\chi= \sum_\alpha e^{- \lambda |\varepsilon_{\alpha}-\varepsilon_{0}|}
\end{equation}
with $\varepsilon_{\alpha}$ being the $\alpha$th eigenvalue,
and $\varepsilon_0$ the ground state~\footnote{For computational purposes, the previous sum can be restricted to a subset of the lowest lying states.}.
The parameter $\lambda$ controls the energy resolution
of the quasi-degeneracy, which in the limiting case
$\lim_{\lambda\rightarrow \infty } \lim_{L\rightarrow \infty } \chi $  becomes the
thermodynamic degeneracy of the ground state~\footnote{We take $1/\lambda (=0.005)$}. We will focus our analysis on system sizes with
$L=16$, that are large enough to show different transition regimes that would converge to the different phases
of the model in the thermodynamic limit.

In addition to the quasi-degeneracy $\chi$, we can characterize the phase boundaries using the electronic
correlation entropy given by~\cite{Gersdorf1997, Huang2006, Esquivel1996, BenavidesRiveros2017, Aikebaier2022}
\begin{align}
    {\cal C}_{\rm corr} = - \frac{1}{L} \sum_{j=1}^{L} s_{j}\log (s_{j}),
\end{align}
where $0 \leq s_{j} \leq 1$ is the $j$th eigenvalue of the correlation matrix. The elements of the correlation matrix $C^{\rm mat}$ are two-point correlation functions that read
$
    C_{ij}^{\rm mat} = | {\rm det} [ \sum_{ll'}^{\ichi} \rho^{ll'}_{ij} ] | \, \, \text{  with }
    \rho_{ij}^{ll'}=\ave{ \Psi_{l} | c^{\dagger}_{i} c_{j}| \Psi_{l'}}
    ,
$
where $\Psi_{l} $ is the $l$th eigenstate on the ground state manifold,
and $\ichi$ is the closest integer to $\chi$. The correlation matrix ${\cal C}_{\rm corr} $ measures many-body entanglement and vanishes in systems described by Hartree-Fock
product states~\cite{Siegbahn1981,PhysRevB.103.235166,Aikebaier2022,2023arXiv230815576S,2023arXiv230607584V}.
It is worth noting that while superconducting states can be represented as a product state in the Nambu basis, the previous definition of correlation entropy yields a finite value for superconducting states.
Large values of ${\cal C}_{\rm corr} $ in certain regions of the phase diagram imply
that the system cannot be represented by a Hartree-Fock product state.

\paragraph*{Machine learning methodology.}
We now present the machine learning methodology to learn
the different regimes of the interacting models, taking as
target functions
$\chi$ and ${\cal C}_{\rm corr}$.
The input of our machine-learning algorithm corresponds
to short-range many-body correlators
in the form of two-point and four-point correlation functions given by
\begin{align}
    d_{ij}&=\ave{c^{\dagger}_{i} c_{j}}_{\ichi}, \quad 
     f_{ij}=\ave{c^{\dagger}_{i} c^{\dagger}_{j}}_{\ichi}, \label{eq:2points} \\
     k_{ij}&=\ave{\kappa_{ij} \kappa^{\dagger}_{ij}}_{\ichi}, \quad 
     p_{ij}=\ave{n_{i} n_{j}}_{\ichi}, \label{eq:4points}
\end{align}
where $\kappa_{ij} = c_{i} c_{j}$ and $\ave{\hat{A } }_{\ichi} \equiv |{\rm det} [\sum_{ll'}^{\ichi} A_{ll'}]|$ with $A_{ll'} = \ave{\Psi_{l} |\hat{A}|\Psi_{l'}}$. Here, $i,j$ run on four neighboring sites in the middle of the chain so that the algorithm relies solely on short-range correlation functions. These correlation functions are used to predict the quasi-degeneracy $\chi$ and the
correlation entropy ${\cal C}_{\rm corr}$.
We collect 20000 different non-Hermitian interacting realizations on the $(U/t,\eta)$ plane,
taking the non-Hermiticity parameter as $\delta \in \{ 0, 0.5\}$.
To predict the quasi-degeneracy, we explore two strategies, the first one is based on transforming the task in a classification
problem for $[\chi]$, and the second one is a regression problem for $\chi$. The prediction of ${\cal C}_{\rm corr}$ is treated as
a regression problem.
The details of our NN architecture for each of these cases are presented in the Supplemental Materials~(SM)~\cite{SuppMat}.

\paragraph*{Results.}

We now present the predictions of different regimes based on various correlators for our Hermitian and non-Hermitian limits. We start with the Hermitian phase diagram shown in Fig.~\ref{fig:chiherm}~(a,b). These panels present the numerical regimes obtained with the exact diagonalization method~\footnote{Numerical calculations are performed using the dmrgpy package in \href{https://github.com/joselado/dmrgpy}{https://github.com/joselado/dmrgpy}. }. The finite-size effect pushed the regime crossovers to smaller $\eta$ values from the
phase boundaries in the thermodynamic limit, a feature that can be systematically analyzed using finite size scaling~\cite{Sayyad2023}.
Performing this scaling gives rise to the thermodynamic phase boundaries shown in the cyan lines~\cite{Sayyad2023}.

The associated predicted regime crossovers using $\chi$ are displayed in Fig.~\ref{fig:chiherm}~(c,d,e,f).
Here, we compare the true (Fig.~\ref{fig:chiherm}ab) and predicted~(Fig.~\ref{fig:chiherm}(c,d,e,f)) phase diagrams obtained from training the NN model using the two-point correlation functions~(Fig.~\ref{fig:chiherm}(c,d)) or the combination of both two-point and four-point correlation functions (Fig.~\ref{fig:chiherm}(e,f)). 
The values of $\ichi$ in Fig.~\ref{fig:chiherm}(a,c,e) are discrete, and the predicted results belong to different classes of $\ichi$. In panels Fig.~\ref{fig:chiherm}(b,d,f), a regression architecture is used to predict $\chi$, and the predicted results~Fig.~\ref{fig:chiherm}(d,f) are obtained as a regression problem.

We now examine how the regimes of the non-Hermitian interacting model can be deduced from short-range correlators using a model trained by the Hermitian dataset with $\delta=0.0$, as shown in Fig.~\ref{fig:chinonherm}. 
Fig.~\ref{fig:chinonherm}(c,d,e,f) shows the predicted phase crossovers
obtained by the algorithm trained with Hermitian data,
which should be compared with true outputs of the non-Hermitian problem shown in Fig.~\ref{fig:chinonherm}(a,b).
Interestingly, the predicted results based on two-point correlation functions 
based on a classification architecture for $\ichi$ (Fig.~\ref{fig:chinonherm}(c)) display a large discrepancy. Such inaccurate prediction is eliminated by incorporating four-point correlation functions into the considered observables, as shown in Fig.~\ref{fig:chinonherm}(e). 
We further note that if we phrase
the task as a regression problem, as shown in Fig.~\ref{fig:chinonherm}(b,d,f), the predicted phase boundaries based on training with two-point correlation functions are more reliable, as shown in Fig.~\ref{fig:chinonherm}(d).  
These results show that the quasidegeneracy of the non-Hermitian model can be extracted from a
model trained purely on Hermitian data.

Aside from $\chi$, the different regimes can be characterized using the correlation entropy ${\cal C}_{\rm corr}$ both in Hermitian $\delta=0$ and non-Hermitian $\delta=0.5t$ systems as respectively shown in Fig.~\ref{fig:corrent}~(a,b). Finite-size effects are reflected in the deviations from the cyan lines,
which are inherited by the changes of $[\chi]$ that impact the definition of the correlation entropy. 
Interestingly, ${\cal C}_{\rm corr} $ exhibits further transitions, quantitatively described by the analytic phase boundaries. The absence of a finite size effect in different regions of the parameter space, 
delineated by the black dashed-dotted lines, signals the exponential convergence towards the ground state due to finite correlation gaps. Similar behavior is reported in Mott insulators~\cite{Jeckelmann2003, Aikebaier2022} and magnetic vortex liquids~\cite{Chern2021}.   
 In Fig.~\ref{fig:corrent}, we present the various regimes for Hermitian (Fig.~\ref{fig:corrent}(a,c,e)) and non-Hermitian (Fig.~\ref{fig:corrent}(b,d,f)) systems using a model trained on Hermitian models with only two-point (Fig.~\ref{fig:corrent}(c,d)) or the combination of two-point and four-point correlation functions (Fig.~\ref{fig:corrent}(e,f)). Overall, all the thermodynamic phase boundaries are
 qualitatively signaled by the correlation entropy. In the non-Hermitian cases, we can identify some regions, mainly inside the black diamond-like phase boundaries, featuring differences from the true results. These differences are reduced 
 when including four-point correlation functions in the training of the Hermitian model; see also the SM~\cite{SuppMat}. 
 It is worth noting that the regions with the most discrepancies have a topological superconducting nature, 
 suggesting that phases with topological and many-body effects require higher-point correlation functions to be inferred with short-range
 information.

Our machine learning models trained only in Hermitian Hamiltonians can characterize the regimes of non-Hermitian interacting systems. 
It is interesting to note that, while we observe a general agreement, small discrepancies between the machine learning predicted regimes
and the computationally exact ones can be observed. This is because non-Hermitian many-body systems can show richer
ground states than their Hermitian analog due to the extent of their spectrum in the complex plane.
As a result, many-body wavefunctions in non-Hermitian models are genuinely different from their Hermitian counterparts,
as these wavefunctions can span different regions of the Hilbert space beyond the original Hermitian training.
Interestingly, this discrepancy opens the possibility of using our machine learning algorithms to directly identify non-Hermitian
phases that do not have a Hermitian counterpart.

\paragraph{Conclusion.}
To summarize, we have demonstrated a transfer machine learning methodology
whereby training on Hermitian many-body models allows us to predict
different regimes of interacting non-Hermitian quantum many-body models. 
This opens the possibility of employing Hermitian many-body physics
to understand the phase boundaries of non-Hermitian systems,
leveraging solutions and methodologies currently only applicable to quantum many-body models.
Our findings reveal that the prediction of
quasi-degeneracy or correlation entropy allows the identification of different regions in interacting systems.
Interestingly, these two methodologies are affected in a qualitatively different manner for finite size effects,
with the correlation entropy showing the fastest convergence to the thermodynamic limit.
Our machine-learning methodology relies on short-range correlation functions, which open the possibility
to potential deployments of our technique in experimental setups.
Our results establish transfer learning as a promising strategy to map regimes on non-Hermitian quantum many-body models
and to identify regimes featuring phenomena not observable in Hermitian models.

\paragraph*{Acknowledgements:}
S.S. thanks F. Marquardt for the helpful discussions.
J.L.L. acknowledges
the computational resources provided by the Aalto Science-IT project,
the financial support from the
Academy of Finland Projects No. 331342, No. 336243 and No 349696,
and the Jane and Aatos Erkko Foundation.

\bibliography{learn_NHintsys}

\newpage
\appendix
\setcounter{secnumdepth}{2}
 \renewcommand{\theparagraph}{\bf \thesubsubsection.\arabic{paragraph}}

\renewcommand{\thefigure}{S\arabic{figure}}
\setcounter{figure}{0} 

\renewcommand{\theequation}{S\arabic{equation}}
\setcounter{equation}{0} 


\section{Details on the architecture of neural networks}

In the main text, we have presented the outcomes of our neural networks~(NN). Here, we provide further details on how we build them up.

The architecture of our neural network~(NN) consists of $1024/1024$ hidden layers when the labels are ${\cal C}_{\rm corr}$. When $\chi$ sets the labels, we consider either a regression problem with $\chi$ being treated as a real number or a classification problem with $\chi$ limited to discrete values of $\{1,2,4\}$ which are analytical orders of degeneracies in different phases. For the regression problem, the NN is built up using  $128/1024/2048/1024/128$ hidden layers while it has $128/1024/3072/1024/128$ hidden layers when the classification problem is explored. The different numbers of hidden layers in NN architecture are attributed to the (dis)continuousness of ${\cal C}_{\rm corr}~(\chi)$. It is clear that the NN for the classification problem trains on more number of parameters. The final layer of the NN is a single dense layer to generate $\chi$ or ${\cal C}_{\rm corr}$. For the classification problem, this last layer has a size of three associated with three classes of $\{1,2,4\}$.

Our NN models are optimized using the Adam algorithm~\cite{Kingma2014} with a learning rate $10^{-6}$. The loss function in the NN architecture  is set to the mean absolute error unless for the classification problem where the categorical cross-entropy is chosen. The validation loss in the training steps is always less than $0.005$. Except for the last layer, we used rectified linear unit~(ReLu) as the activation function. The last layer for training data has the softmax activation function for the classification problem and is set to linear activation function in other cases. 

\section{Comparing predicted and true phase diagrams}
\begin{figure*}
    \centering
    \includegraphics[width=1.9\columnwidth]{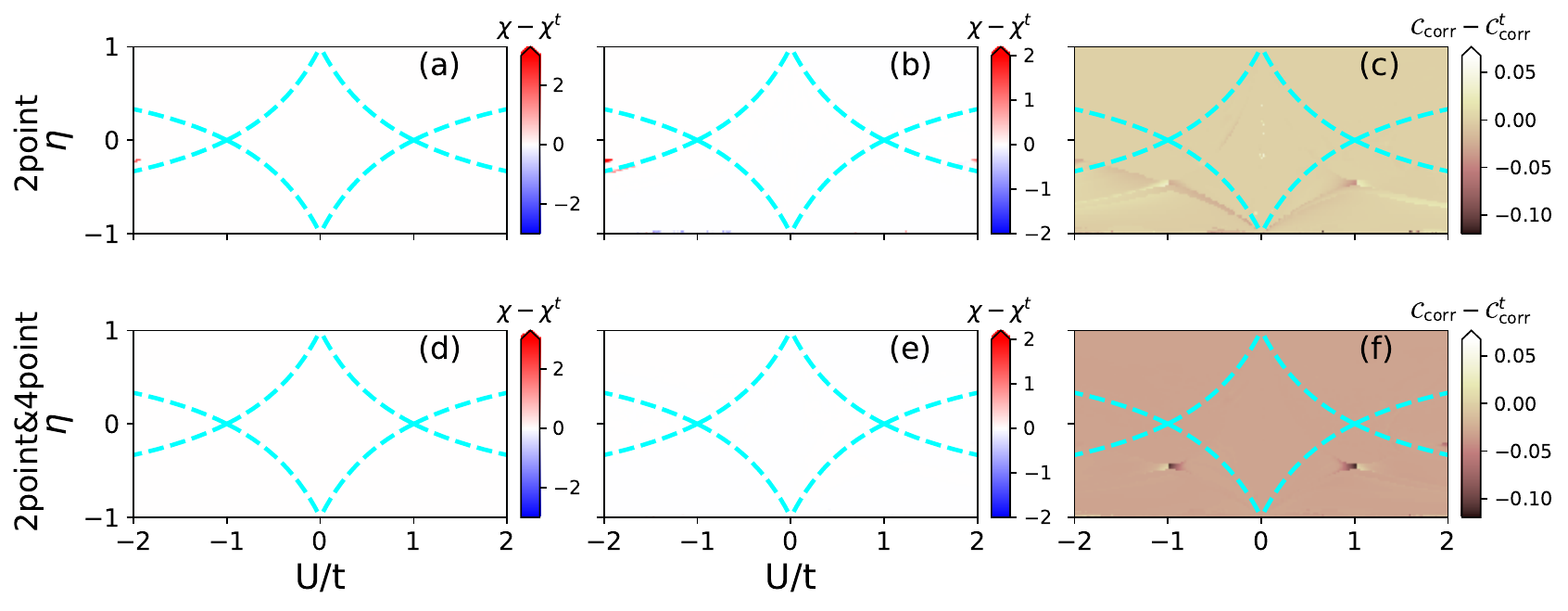}
    \caption{The difference between the predicted phase diagrams trained on the Hermitian dataset with the true Hermitian phase diagram at $\delta=0.0$. The color bars denote the difference between the true~(indicated by the superscript $t$) and predicted values of $\chi$ in panels (a), (b), (d), and (e) or ${\cal C}_{\rm corr}$ in panels (c) and (f). 
    The top panels are obtained for predicated phase diagrams using only two-point correlation functions shown in Fig.~\ref{fig:chiherm} (c) and (d). The bottom panels display the difference between the true and predicted values using both two-point and four-point correlation functions, shown in Fig.~\ref{fig:chiherm} (e), and (f).
    }
    \label{fig:diff-delta0}
\end{figure*}

To better understand the performance of the employed learning method, we compare the predicted and true phase diagrams. We note that the predicted phase diagram for the Hermitian model in Fig.~\ref{fig:chiherm} agrees quantitatively with the true phase diagram as shown in Fig.~\ref{fig:diff-delta0}. This quantitative agreement between the predicted and true numerical results remains intact even when the non-Hermiticity is finite; see Sec.~\ref{sec:pred_nhdata_nhmodel}. This suggests that the capability of NN in learning many-body effects is not limited to the realm of Hermitian physics.

\begin{figure*}
    \centering
    \includegraphics[width=1.9\columnwidth]{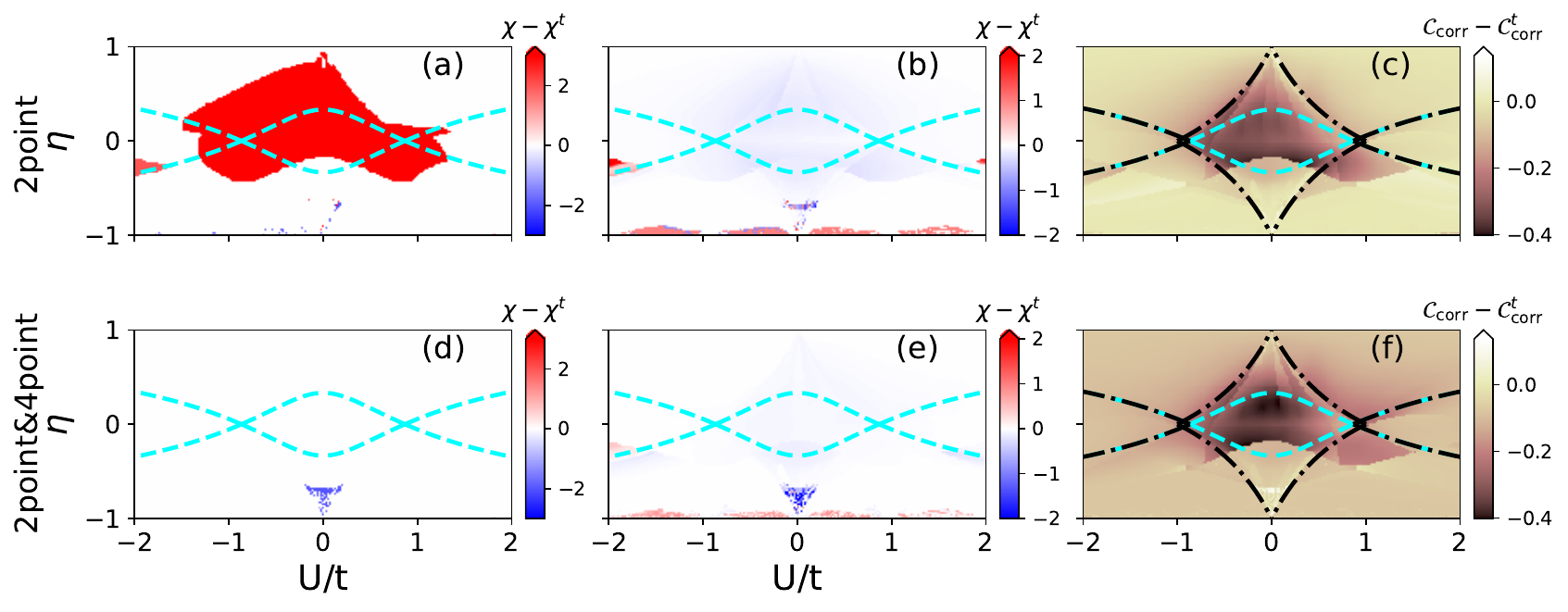}
    \caption{The difference between the predicted phase diagrams with $\delta=0.5t$ trained on the Hermitian dataset with the true Hermitian phase diagram at $\delta=\delta_{h}=0.0$. The color bars denote the difference between the true~(indicated by the superscript $t$) and predicted values of $\chi$ in panels (a), (b), (d), and (e) or ${\cal C}_{\rm corr}$ in panels (c) and (f). 
    The top panels are obtained for predicated phase diagrams using only two-point correlation functions shown in Fig.~\ref{fig:corrent} (b), (e), and (h). The bottom panels display the difference between the true and predicted values using both two-point and four-point correlation functions, shown in Fig.~\ref{fig:corrent} (c), (f), and (i).}
    \label{fig:diff-delta05}
\end{figure*}

We, however, note that the differences between true and predicted phase diagrams of the non-Hermitian case, obtained using models on the Hermitian dataset display more features. 
Figure~\ref{fig:diff-delta05} (a) and (d) display the difference between the true~(indicated by superscript $t$) and predicted phase diagrams using values of $\chi$ as the classifier. It is evident that a large portion of the phase diagram is incorrectly characterized where only two-point correlation functions are used to train the model; see panel (a). By employing both two-point and four-point correlation functions, the predicted phase diagram is significantly improved and merely small differences along the phase boundary and around $U=0$ remain. 

If we train the NN model using continuous values of $\chi$ as shown in panels (b) and (e), the predicted values using only two-point correlation functions are more accurate than in panel (a). Here, we again witness some discrepancy from the true results around $U=0$, $U\approx \pm 2t$, or around $\eta=-1$. The results are slightly modified when four-point correlation functions are also used in training the NN model as inaccuracies around large values of $U$ are eliminated.

When we employ ${\cal C}_{\rm corr}$ instead of $\chi$, the predicted results in the absence~(c) or presence~(f) of four-point correlation functions exhibit similar phase diagrams; although in difference is less pronounced in the absence of four-point correlation functions in panel (c). This might be rooted in the nature of the correlation entropy which is constructed from two-point correlation functions.

\section{Predicting non-Hermitian phase diagrams using non-Hermitian dataset}\label{sec:pred_nhdata_nhmodel}
\begin{figure*}[t!]
    \centering
    \includegraphics[width=1.9\columnwidth]{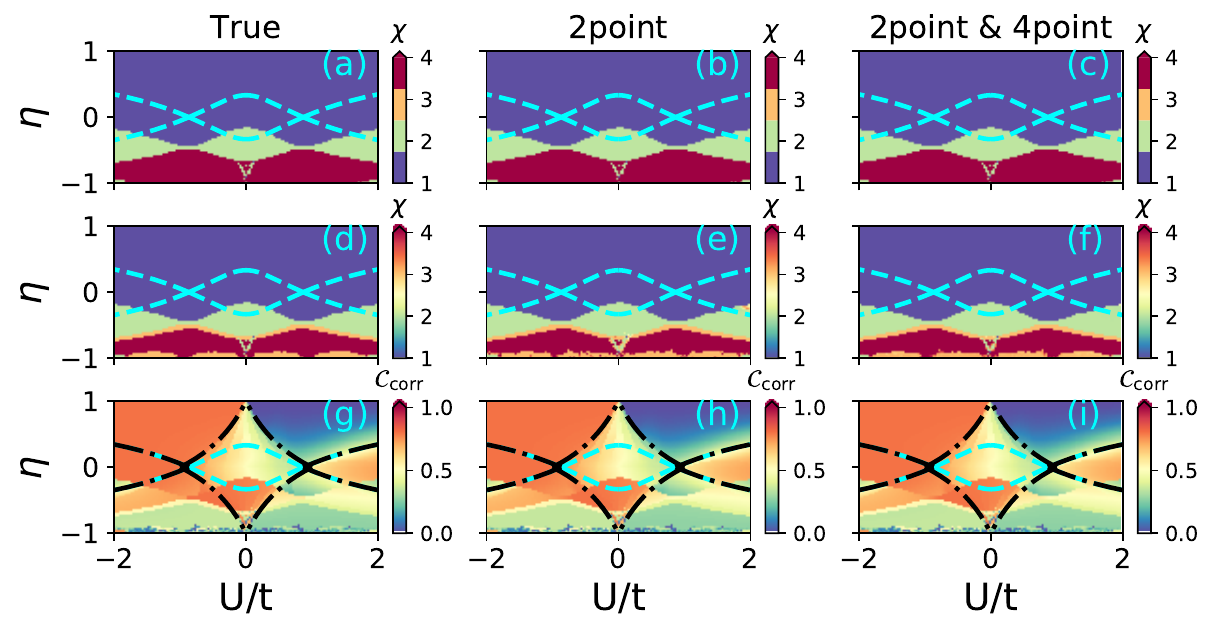}
    \caption{
    The phase diagram of the Kitaev-Hubbard chain with $L=16$ on the $U/t-\eta$ plane at $\delta=0.5t$. The color bar denotes $\chi$~(top and middle panels) and ${\cal C}_{\rm corr}$~(bottom). The results in (a), (d), and (g) are calculated by exact diagonalization. The phase diagrams on the middle~(right) column are obtained using the NN model trained by two-point~(both two-point and four-point) correlation functions. $\ichi$ in the top row is treated as a discrete classifier with $\ichi \in \{ 1,2,4\}$ while it is treated as a continuous number in the panels of the middle row.
    }
    \label{fig:corrent_deltam05}
\end{figure*}

\begin{figure*}
    \centering
    \includegraphics[width=1.9\columnwidth]{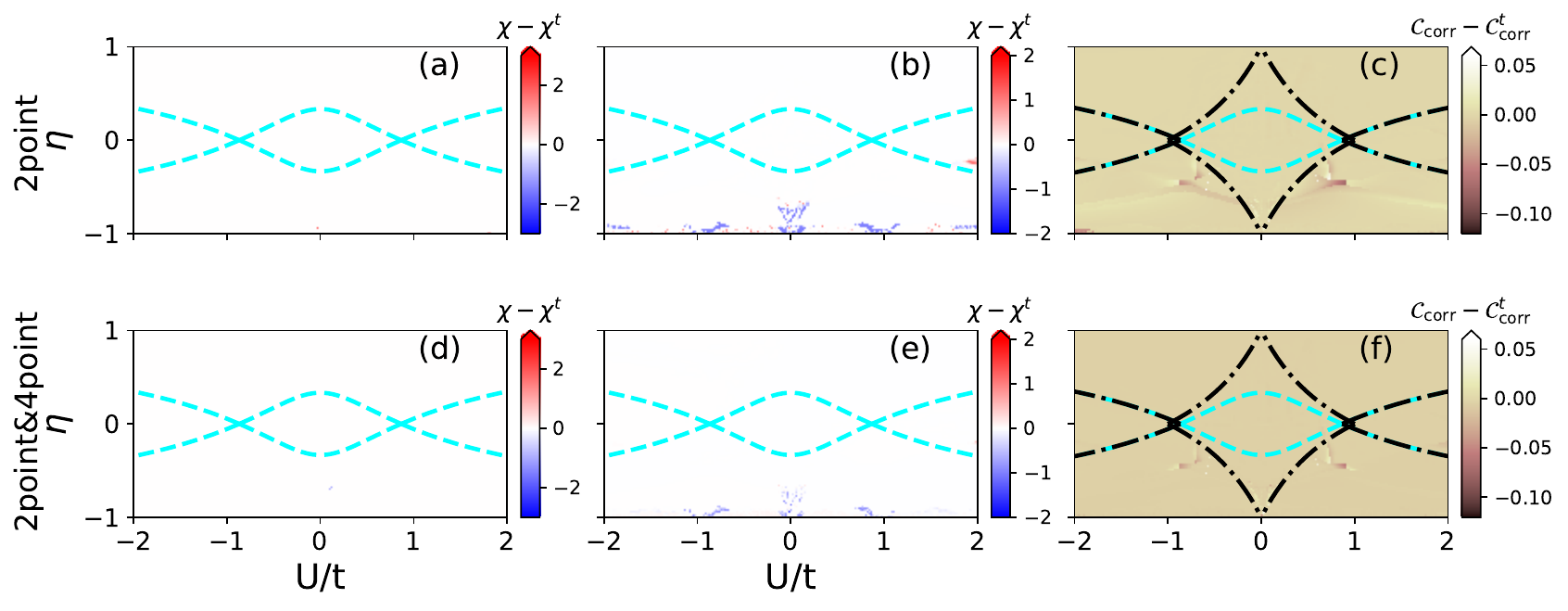}
    \caption{The difference between the predicted phase diagrams trained on the non-Hermitian dataset with the true non-Hermitian phase diagram at $\delta=0.5t$. The color bars denote the difference between the true~(indicated by the superscript $t$) and predicted values of $\chi$ in panels (a), (b), (d), and (e) or ${\cal C}_{\rm corr}$ in panels (c) and (f). 
    The top panels are obtained for predicated phase diagrams using only two-point correlation functions shown in Fig.~\ref{fig:corrent_deltam05} (b), (e), and (h). The bottom panels display the difference between the true and predicted values using both two-point and four-point correlation functions, shown in Fig.~\ref{fig:corrent_deltam05} (c), (f), and (i).
    }
    \label{fig:diff-delta05-deltam05}
\end{figure*}

In the main text, we present the prediction of non-Hermitian phase diagrams using NN models trained on Hermitian datasets. There, we have detected some discrepancies which we have attributed to the properties of non-Hermitian wavefunctions. Here, we present the prediction of non-Hermitian phase diagrams using NN models trained on the non-Hermitian datasets with $\delta=0.5t$. Figure \ref{fig:corrent_deltam05} displays the true~(left column) and predicted~(middle and left columns) phase diagrams. The NN models are trained using two-point correlation functions~(middle column) or the combination of two-point and four-point correlation functions~(right column). The employed labels for the data sets are the discrete values of $\chi$~(top row), the numerical values of $\chi$~(middle), and ${\cal C}_{\rm corr}$~(bottom). In all panels, the phase diagrams are quantitatively reproduced, as the differences between the true~(indicated by superscript $t$) and predicted values are very tiny as shown in Fig.~\ref{fig:diff-delta05-deltam05}.

\end{document}